# High frequency magnetic behavior through the magnetoimpedance effect in CoFeB/(Ta, Ag, Cu) multilayered ferromagnetic thin films

M. S. Marques[1], T. Mori[2], L. F. Schelp[2], C. Chesman[1], F. Bohn[3], and M. A. Corrêa[1]*

[1]*Departamento de Física Teórica e Experimental, Universidade Federal do Rio Grande do Norte, 59072-970 Natal, RN, Brazil*

[2]*Departamento de Física, Universidade Federal de Santa Maria, 97105-900 Santa Maria, RS, Brazil*

[3]*Escola de Ciências e Tecnologia, Universidade Federal do Rio Grande do Norte, 59072-970 Natal, RN, Brazil*

**Abstract**

We studied the dynamics of magnetization through an investigation of the magnetoimpedance effect in CoFeB/(Ta, Ag, Cu) multilayered thin films grown by magnetron sputtering. Impedance measurements were analyzed in terms of the mechanisms responsible for their variations at different frequency intervals and the magnetic and structural properties of the multilayers. Analysis of the mechanisms responsible for magnetoimpedance according to frequency and external magnetic field showed that for the CoFeB/Cu multilayer, ferromagnetic resonance (FMR) contributes significantly to the magnetoimpedance effect at frequencies close to 470 MHz. This frequency is low when compared to the results obtained for CoFeB/Ta and CoFeB/Ag multilayers and is a result of the anisotropy distribution and non-formation of regular bilayers in this sample. The $MI_{max}$ values occurred at different frequencies according to the used non-magnetic metal. Variations between 25% and 30% were seen for a localized frequency band, as in the case of CoFeB/Ta and CoFeB/Ag, as well as for a wide frequency range, in the case of CoFeB/Cu.

*Corresponding author. Tel: +55 84 36068089; fax: +55 84 32119217.
E-mail address: marciocorrea@dfte.ufrn.br (M. A. Corrêa)

## 1. Introduction

The magnetoimpedance (MI) effect has attracted considerable interest in recent times mainly due to its potential application in high-performance sensors and sensing elements in integrated circuits, which can be used in a wide variety of sectors, such as biology and geo-studies [1, 2].

MI effect is related to a change in impedance in a sample when subjected to an external magnetic field. At moderate and high frequencies of tens of MHz to GHz, the main mechanisms responsible for variation in impedance are the strong skin effect, which affects transversal permeability, and the ferromagnetic resonance (FMR) effect, which occurs due to the configuration of magnetic fields in the sample.

This magnetoimpedance effect has been studied extensively in samples in the form of ribbons [3, 4, 5] and microwires [6, 7, 8]; however, more recently, there has been increased interest in the effect of single and multilayered thin films [9, 10, 11, 12, 13, 14, 15], which have interesting magnetic and structural properties, that can be used in integrated devices sensors.

In the specific case of thin films, samples of F/NM multilayers, where F is a ferromagnetic alloy and NM is a non-magnetic metallic material such as Cu, Ag or Ta [14, 15], generally have excellent properties for the study of the MI effect. Among the properties which stand out are high magnetic permeability, high values for saturation magnetization, low relaxation parameter values ($\alpha$) and high electrical conductivity

In this spirit, for some years now, our group has carried out an intensive search for different ferromagnetic alloys for the growth of multilayers [14, 15, 16]. CoFe alloys have great technological potential, being used, for example, in applications such as microwave absorbers [17], shielding materials [18], strain gauges [19] and magnetic random access memories (MRAMs) [20]. For this reason, we carried out a study of the magnetic and structural properties, as well as the MI effect, of multilayers where the ferromagnetic material is a CoFeB alloy and the non-magnetic metallic (NM) spacer material is Ta, Ag and Cu. In particular, we present the results for MI effect obtained in multilayers produced with different spacers and we discuss these results in terms of the magnetic and structural properties of the multilayers.

## 2. Experiment

In this study, we produced a series of multilayers ($Co_{40}Fe_{40}B_{20}$/(NM)) × 50, where NM = Ta, Ag or Cu), grown using the magnetron sputtering technique. The samples were deposited onto glass substrates, covered by a Ta buffer layer, and finally covered again by a Ta cap layer. Fig. 1 shows the structure of the samples deposited. In these samples, deposition time was necessary to obtain the nominal thickness of the 9 nm ferromagnetic layers, 2 nm non-magnetic layers, and 5 nm buffer and cap layers. The deposition process was carried out according to the following parameters: base pressure of $4.5 \times 10^{-8}$ Torr, Ar pressure during deposition of $2.0 \times 10^{-3}$ Torr, 90 W RF power supply for the CoFeB alloy and a DC current of 25 mA for the Ta, Ag or Cu metallic layers. Using these parameters, deposition rates were 0.33 nm/s for $Co_{40}Fe_{40}B_{20}$ and 0.18 nm/s, 0.91 nm/s and 0.29 nm/s for Ta, Ag and Cu, respectively.

In order to induce magnetic anisotropy and define an easy magnetization axis, the substrate were submitted to a 1 kOe magnetic field inside the chamber during deposition. Fig. 1 shows the direction of the magnetic field, indicated by the arrow, applied transversely to the main axis of the sample. Magnetic properties were obtained using an alternate gradient field magnetometer (AGM), at room temperature with a maximum field of ±300 Oe, in two directions: along the direction defined by the magnetic field applied during deposition (EA), and perpendicular to this direction (HA).

The structural properties of the multilayers were initially obtained from measurements of X-ray reflectivity (XRR). The diffractograms were obtained at small angles $\theta$ from 0.25º to 5.00º, using a Broker AXS D8 Advance diffractometer, and Bragg-Brentano geometry ($\theta$-$2\theta$) with Göbel mirrors at the outlet of the tube in order to make the beam parallel. To complete the structural analysis of the samples, measurements of the DC electrical transport in the films were taken using a 4-point probe.

To measure impedance, samples were cut to a size of 3 x 10 mm$^2$, with the main axis in HA direction, fixed to a stripline type sample holder using silver glue with a 24-hr curing time. Impedance was measured using an HP4396B spectrum-impedance-network analyzer equipped with an HP43961A impedance measuring fixture and a microstrip line cavity. Detailed information on this procedure can be found in [21]. Measurements were obtained from a field of ±300 Oe over a wide frequency range, from 10 MHz to 1.8 GHz. The MI ratio was defined as

$$MI\% = \frac{Z(H) - Z(H_{max})}{Z(H_{max})}, \quad (1)$$

where $Z(H)$ is the impedance value at a given field $H$ and $Z(H_{max})$ is the impedance at the maximum applied magnetic field $H_{max}$, where the magnetic state of the sample is saturated. Finally, $MI_{max}$ is defined as the maximum $MI\%$ value for a given frequency.

## 3. Results and discussion

### 3.1. Structural properties

The XRR measurements for CoFeB/Ta, CoFeB/Ag and CoFeCu can be found in items (a) (b) and (c) of Fig. 2, respectively. This figure also includes simulations performed from ideal multilayers with perfect interfaces. Due to the presence of Bragg peaks, all the samples exhibit a chemical modulation perpendicular to the substrate, with periods of 11 nm for the CoFeB/Ta and CoFeB/Ag samples and 12 nm for the CoFeB/Cu sample. X-ray patterns also indicate that the morphology of the multilayers differs from one sample to another. The quality of the multilayer (i.e. layer thickness, homogeneity and interface flatness) depends substantially on the difference in surface free energy between the materials included in the sample. When Ta is used as magnetic material spacer, this difference is small and the growth mode leads to layers with almost constant thickness and flat interfaces. As can be seen, the relative intensity of the successive peaks decreases slowly and is similar to that found in a simulation of an ideal multilayer. Ag and Cu, on the other hand, are immiscible with Co and Fe and tend to have very different surface-free energies from those of magnetic materials. They do not wet the CoFeB surface, generating layers with less homogeneous thickness and rougher interfaces. This growth mode is reflected in the diffractogram as a rapid decrease in the intensity of successive Bragg peaks, departing from the pattern that would be expected from a perfect multilayer. Above a certain level, these fluctuations in thickness along the layers may even destroy the correct material stacking in parts of the samples. A comparison of the electrical resistivity in the three samples suggests that this may occur in the CoFeB/Cu sample, since it shows greater resistivity than that found in CoFeB/Ag. This result is inconsistent with the presence of a parallel arrangement of layers. The structural characteristics, confirmed in the XRR measurements, led to significant distortions in their magnetic properties.

**3.2 Quasi-static magnetic properties**

The magnetic properties of the multilayers were measured through magnetization curves obtained from a magnetic field oriented along and transversal to the direction of the magnetic field applied during deposition. Fig. 3 shows the magnetization curves measured for our set of samples.

By comparing these curves, it is possible to verify the magnetic anisotropy induced during film growth for the CoFeB/Ta and CoFeB/Ag samples, confirming the easy magnetization axis EA oriented transversely to the main axis of the sample and along the same direction of the magnetic field applied during deposition. In these cases, saturation fields are approximately 70 Oe and 46 Oe for CoFeB/Ta and CoFeB/Ag samples, respectively.

On the other hand, in the CoFeB/Cu sample, we perceived weak anisotropy induction and an increase in the coercive field when compared to results for CoFeB/Ta and CoFeB/Ag samples. In particular, this fact is associated with the non-formation of regular bilayers in the CoFeB/Cu multilayer, as previously discussed. The formation of islands instead of continuous layers results in an increase in structural disorder and subsequent storage of higher internal stress as the sample is grown; this gives rise to random local-field anisotropies which result in weak effective anisotropy oriented in the direction of the magnetic field applied during deposition.

As effective anisotropy directly influences the dynamic results of magnetization, the MI effect measurements of these multilayers reflect exactly the differences discussed.

**3.3 MI effect results**

Fig. 4 shows impedance $Z$ as a function of magnetic field and frequency for the CoFeB/Ta, CoFeB/Ag and CoFeB/Cu samples. While all the impedance measurements in this study were obtained from a wide range of frequency levels (10 MHz to 1.8 GHz), a very low signal-to-noise ratio was recorded for frequencies up to 300 MHz. This result is related to the low thickness of the multilayers grown and the subsequent appearance of the skin effect only at higher frequency levels. For this reason, $Z$ behavior is only shown for frequencies of 500MHz or higher.

For the whole frequency range, all samples exhibit similar behavior, characterized by a double peak structure, a feature obtained when the magnetic field and the current are transversally applied to the easy magnetization axis [3].

Fig. 4 shows that in the CoFeB/Ta sample, maximum $Z$ position remains stable up to a frequency of ~990 MHz, while in the CoFeB/Ag sample, the change in $Z$ peak position occurs at frequencies higher than ~1.19 GHz. This result indicates that the behavior of both samples is very similar. Nonetheless, the most striking result can be seen in the CoFeB/Cu sample, where a displacement in $Z$ peaks was observed at much lower frequency levels, from ~450 MHz.

In all samples, along the entire frequency band in which $Z$ peak position remains stable, the main mechanism responsible for MI behavior is the strong skin effect. On the other hand, the change in $Z$ peak position in accordance with magnetic field as frequency increases reflects the fact that FMR effect becomes the main mechanism responsible for variations in impedance. In all samples, the contribution of the FMR effect to $Z$ was verified using the method described by Barandiarán *et al.* [22], and previously employed by our group in [14]. In this case, in the CoFeB/Ta and CoFeB/Ag samples, the contribution of FMR effect to MI behavior was observed at frequency levels starting at ~760 MHz and ~950 MHz, respectively. In the CoFeB/Cu sample, a small contribution of FMR to the MI effect was observed starting at frequency levels of ~420 MHz. This behavior at a lower frequency level may be associated with the presence of random local anisotropies induced by the formation of islands instead of continuous

Cu layers, and the subsequent increase in stored internal stress. This favors the appearance of localized resonances, contributing to MI variation starting at lower frequency levels.

Finally, Fig. 5 shows $MI_{max}$ vs. $f$ behavior obtained from measurements of impedance in our multilayers. A single peak close to 1 GHz with variation of around 30%, associated with FMR, can be seen in the CoFeB/Ta and CoFeB/Ag samples. This behavior is similar to that obtained by our group in multilayers produced with other compositions [14, 21, 23].

The same figure also demonstrates the influence of the FMR effect in the CoFeB/Cu sample. In this case, the distribution of stress-induced anisotropies and the non-formation of complete Cu layers results in smaller impedance variations of around 25%. However, these variations occur for a wide range of frequencies.

Thus, this new framework qualifies samples with distinct anisotropies for different technological applications. While samples with well-defined anisotropies are characterized by greater impedance variations within a narrow frequency band and can be used in devices requiring localized frequency response with high sensitivity, samples with weakly induced anisotropies, when used as sensor elements, are able to provide sensitivity which, albeit lower, is almost linear even with frequency variations in the excitation circuit during use.

## 4. Conclusion

In this study, we presented the structural and magnetic (static and dynamic) properties of CoFeB/(Ta, Ag, Cu) multilayers The structural properties obtained from measurements of XRR in the CoFeB/Ta and CoFeB/Ag samples were characterized by good bilayer uniformity when compared to those of the CoFeB/Cu sample, whose structure consists of CoFeB layers interspersed with Cu islands. Such structural properties influenced the magnetization behavior, as can be verified through the effective anisotropies observed in the magnetization curves. The mechanisms responsible for greater variations in MI effect occur at high frequencies associated with the emergence of FMR effect in the CoFeB/Ta and CoFeB/Ag samples. On the other hand, in the CoFeBCu sample, due to the formation of islands in the Cu layer and the weak induction of effective anisotropy resulting from localized anisotropies oriented in different directions, we observed the contributions of FMR at much lower frequency levels than in the other multilayers.

With these results and those previously published by our group [14, 21, 23], we end a cycle of studies related to the MI effect in multilayers with different ferromagnetic alloys. Here, we confirmed variations of up to 30% in the CoFeB/Ta sample at a frequency level of 900 MHz, and variations of 27% at a frequency level of 1.2 GH in the CoFeB/Ag sample. Based on these results, it can be inferred that the proper choice of the ferromagnetic alloy allows tuning the largest percentage changes in MI to a given frequency. This fact is important for the use of these alloys as sensor elements at certain frequencies. Finally, the CoFeB/Cu sample, in particular, demonstrated behavior relevant to the use of sensor elements requiring constant variations in the MI effect within a wide frequency range. In this sample, we noted an efficiency level of approximately 25%, obtained, however, within a wide range of frequencies from 600 MHz to 1 GHz.

**Acknowledgments**

The authors were supported by CNPq, CAPES, FAPERJ, FAPERGS and FAPERN.

**Figure captions**

Fig. 1. Schematic structure of the multilayer. The sample consists of a multilayer, formed by 50 bi-layers of CoFeB/NM, where NM = Ta, Ag or Cu are deposited onto a glass substrate covered by a Ta buffer layer and, finally, covered by a Ta cap layer. The larger arrow indicates the direction of the magnetic field applied during sample deposition, transversely to the main axis of the sample.

Fig. 2. XRR experimental patterns (black lines) for our multilayers, together with theoretical simulations (red lines), calculated for model structures, without any degree of roughness and/or interdiffusion at the interfaces, and layer thicknesses of 9 nm for CoFeB and 2 nm for Ag or Cu. Calculations were made using the "Software for modeling the optical properties of multilayer films – IMD", whose methods are outlined in detail in [24].

Fig. 3. Magnetization curves for the CoFeB/Ta, CoFeB/Ag and CoFeB/Cu samples, obtained when the magnetic field is applied along (open circles) and transversely (closed squares) to the direction defined by the magnetic field applied during the deposition, i. e., perpendicularly to the main axis of the sample.

Fig. 4. Impedance $Z$ as a function of the magnetic field and frequency for the CoFeB/Ta (top), CoFeB/Ag (middle) and CoFeB/Cu (bottom) samples. All the $Z$ curves exhibit hysteretic behavior; however, to clarify the general behavior, only the curves obtained when the field goes from the negative to the maximum positive values are presented.

Fig. 5. $MI_{max}$ versus $f$ for the CoFeB/Ta, CoFeB/Ag and CoFeB/Cu samples.

**Figures**

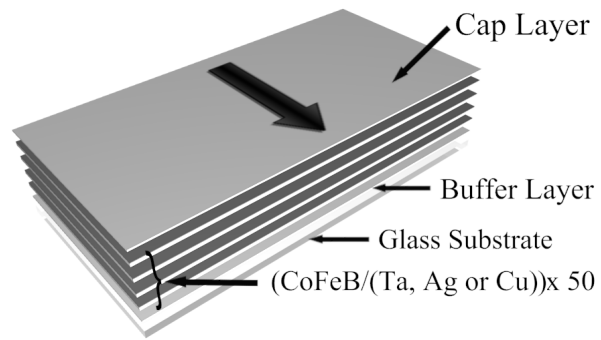

Fig. 1. M. S. Marques *et al.*, reduction factor: 1.0.

10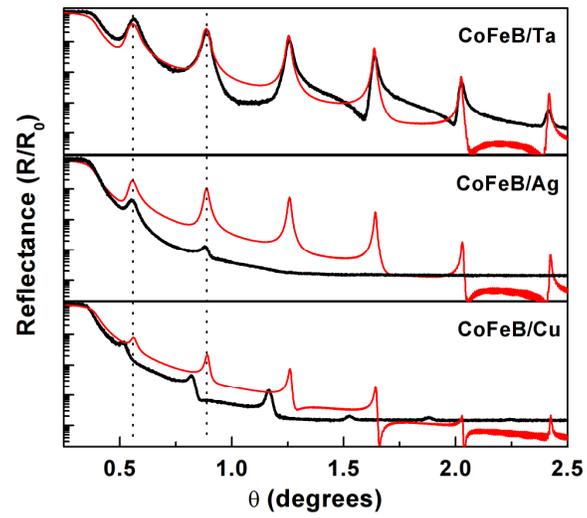

Fig. 2. M. S. Marques *et al.*, reduction factor: 1.0.

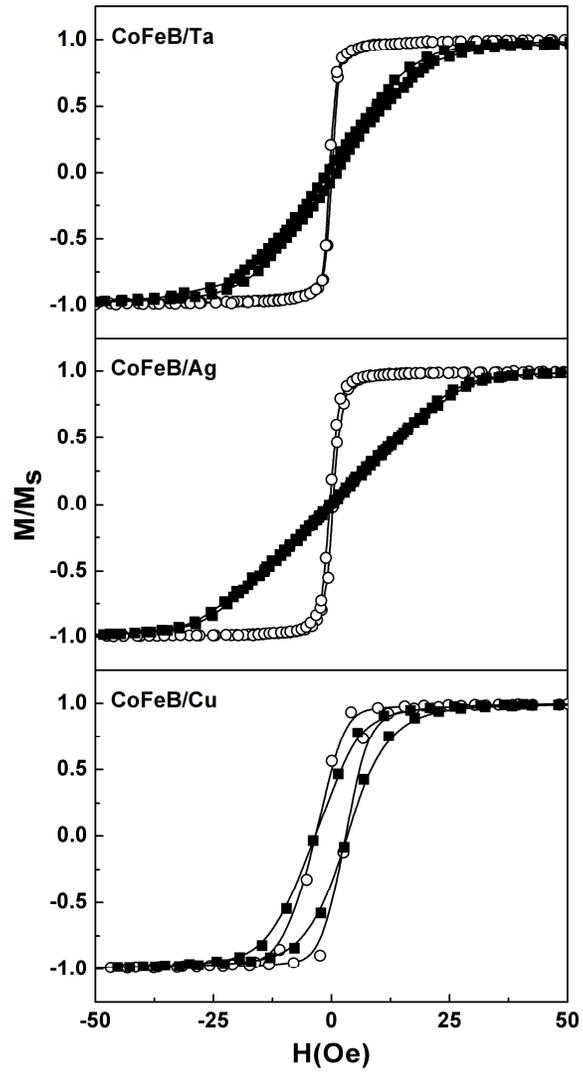

Fig. 3. M. S. Marques *et al.*, reduction factor: 1.0.

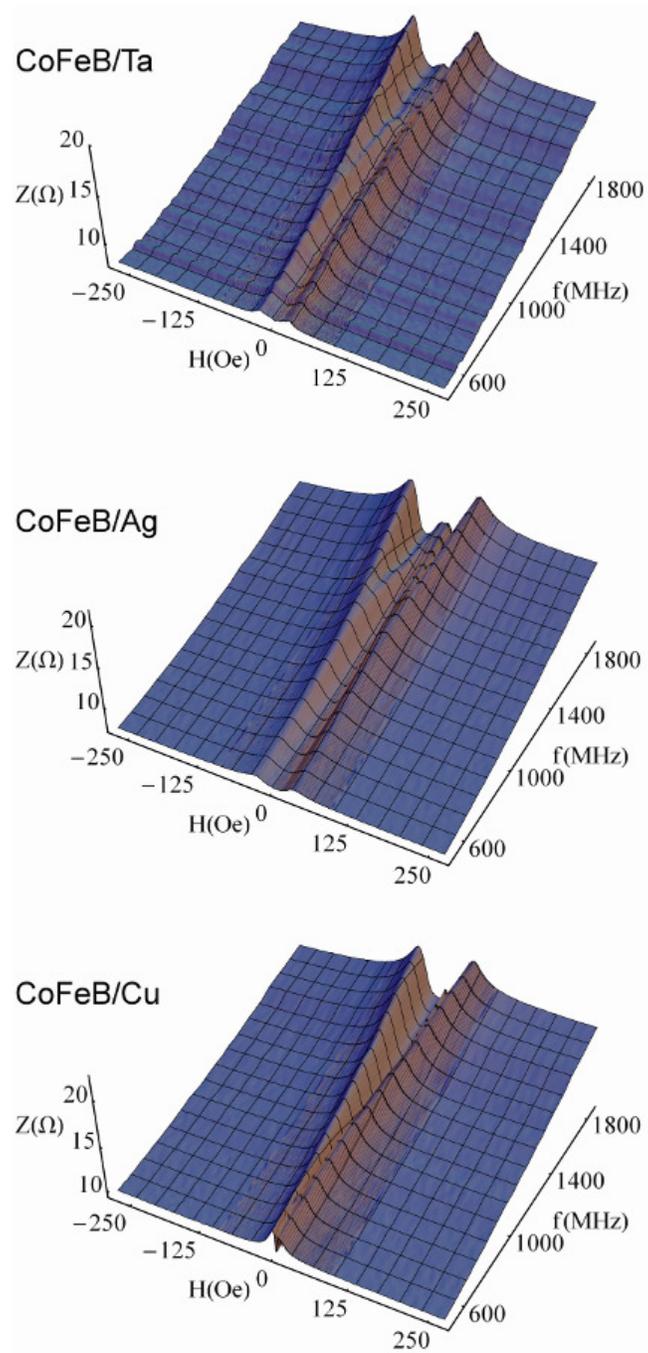

Fig. 4. M. S. Marques *et al.*, reduction factor: 1.0.

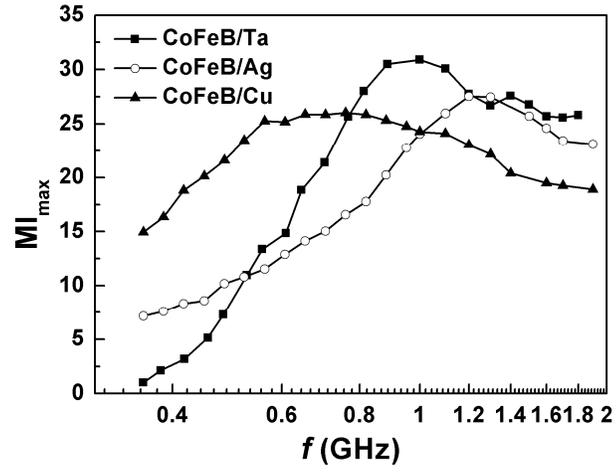

Fig. 5. M. S. Marques *et al.*, reduction factor: 1.0.